# On Spacetime Structure and Electrodynamics*

Wei-Tou Ni

*School of Optical-Electrical and Computer Engineering,
University of Shanghai for Science and Technology,
516, Jun Gong Rd., Shanghai 200093, China* weitouni@163.com

*Center for Gravitation and Cosmology,
Department of Physics, National Tsing Hua University,
No. 101, Kuang Fu II Rd., Hsinchu, Taiwan, ROC 30013*
weitou@gmail.com

Electrodynamics is the most tested fundamental physical theory. Relativity arose from the completion of Maxwell-Lorentz electrodynamics. Introducing the metric $g_{ij}$ as gravitational potential in 1913, versed in general (coordinate-)covariant formalism in 1914 and shortly after the completeion of general relativity, Einstein put the Maxwell equations in general covariant form with only the constitutive relation between the excitation and the field dependent on and connected by the metric in 1916. Further clarification and developments by Weyl in 1918, Murnaghan in 1921, Kottler in 1922 and Cartan in 1923 together with the corresponding developments in electrodynamics of continuous media by Bateman in 1910, Tamm in 1924, Laue in 1952 and Post in 1962 established the premetric formalism of electrodynamics. Since almost all phenomena electrodynamics deal with have energy scales much lower than the Higgs mass energy and intermediate boson energy, electrodynamics of continuous media should be applicable and the constitutive relation of spacetime/vacuum should be local and linear. *What is the key characteristic of the spacetime/vacuum?* It is the Weak Equivalence Principle (WEP I) for photons/wave packets of light which states that the spacetime trajectory of light in a gravitational field depends only on its initial position and direction of propagation, and does not depend on its frequency (energy) and polarization, i.e. nonbirefringence of light propagation in spacetime/vacuum. With this principle it is proved by the author in 1981 in the weak field limit, and by Lammerzahl and Hehl in 2004 together with Favaro and Bergamin in 2011 without assuming the weak-field condition that the constitutive tensor must be of the core metric form with only two additional degrees of freedom – the pseudoscalar (Abelian axion or EM axion) degree of freedom and the scalar (dilaton) degree of freedom (i.e. metric with axion and dilaton). In this paper, we review this connection and the ultrahigh precision empirical tests of nonbirefringence together with present status of tests of cosmic Abelian axion and dilaton. If the stronger version of WEP is assumed, i.e. *WEP II for photons* (wave packets of light) which states in addition to WEP I also that the polarization state of the light would not change (e.g. no polarization rotation for linear polarized light) and no amplification/attenuation of light, then no Abelian (EM) axion and no dilaton, and we have a pure metric theory.

---

# 1. Introduction

In this exposition, we give an overview on basics, core derivations and empirical foundations of spacetime structure and electrodynamics to serve as an introduction to the current special issue on Spacetime Structure and Electrodynamics. Immediately folloing this overview in this issue is a review "On Kottler's path: origin and evolution of the premetric program in gravity and electrodynamics" of Hehl, Itin and Obukhov [1]. The two articles form complements of each other.

Relativity arose from Maxwell-Lorentz theory of electromagnetism. Maxwell equations in Gaussian units are

$$\nabla \cdot \boldsymbol{D} = 4\pi \rho, \tag{1a}$$
$$\nabla \times \boldsymbol{H} - \partial \boldsymbol{D}/\partial t = 4\pi \boldsymbol{J}, \tag{1b}$$
$$\nabla \cdot \boldsymbol{B} = 0, \tag{1c}$$
$$\nabla \times \boldsymbol{E} + \partial \boldsymbol{B}/\partial t = 0, \tag{1d}$$

where $\boldsymbol{D}$ is the displacement, $\boldsymbol{H}$ the magnetic field, $\boldsymbol{B}$ the magnetic induction, $\boldsymbol{E}$ the electric field, $\rho$ the electric charge density, and $\boldsymbol{J}$ the electric current density (See, e.g., Jackson [2], p. 218 (6.28)). We use units with the nominal light velocity $c$ equal to 1. With the sources known, from these equations with 8 components we are supposed to be able to solve for the unknown fields $\boldsymbol{D}$, $\boldsymbol{H}$, $\boldsymbol{B}$ and $\boldsymbol{E}$ with 12 degrees of freedom. These equations form an under determined system unless we supplement them with relations. The relations are the constitutive relation between ($\boldsymbol{D}$, $\boldsymbol{H}$) and ($\boldsymbol{E}$, $\boldsymbol{B}$) [or ($\boldsymbol{D}$, $\boldsymbol{B}$) and ($\boldsymbol{E}$, $\boldsymbol{H}$)]:

$$(\boldsymbol{D}, \boldsymbol{H}) = \chi(\boldsymbol{E}, \boldsymbol{B}), \tag{2}$$

where $\chi(\boldsymbol{E}, \boldsymbol{B})$ is a 6-component functional of $\boldsymbol{E}$ and $\boldsymbol{B}$. With the constitutive relation, the unknown degrees of freedom become 6, the Maxwell equations seem to be over determined. Note that if we take the divergence of (1d), by (1c) it is automatically satisfied. Hence (1c) and (1d) (the Faraday tetrad) have only 3 independent equations. If we take the divergence of (1b), by (1a) it becomes the continuity equation

$$\nabla \cdot \boldsymbol{J} + \partial \rho/\partial t = 0, \tag{3}$$

a constraint equation on the sources. Hence, (1a) and (1b) (the Ampère-Maxwell tetrad) have only 3 independent equations also. To form a complete system of equations, we need equations governing the action of the electric field and magnetic induction on the charge and current. Lorentz force law provides this link and completes the system:

$$\boldsymbol{F} = m \, d\boldsymbol{v}/dt = q \, (\boldsymbol{E} + \boldsymbol{v} \times \boldsymbol{B}), \tag{4}$$

where $\boldsymbol{v}$ is the velocity of the charge and $\boldsymbol{F}$ is the force on it due to electric field and magnetic induction.

On 21 December 1907, Minkowski read before the Academy "Die Grundgleichungen für die elektromagnetischen Vorgänge in bewegten Körpern" (The fundamental equations for electromagnetic processes in Moving bodies) [3]. In this paper, Minkowski put Maxwell equations into geometric form in four-dimensional spacetime with Lorentz covariance using Cartesian coordinates $x$, $y$, $z$ and imaginary time $it$ and numbering them as $x_1 \equiv x$, $x_2 \equiv y$, $x_3 \equiv z$ and $x_4 \equiv it$. Minkowski defined the 4-dim excitation in terms of **D** and **H**, and the 4-dim field strength in terms of **E** and **B**.

Maxwell equations in Minkowski form was soon written in integral form by Hargreaves [4] and devoted a detailed investigation by Bateman [5] and Kottler [6].

In 1909, Bateman [5] worked on the electrodynamical equations. He used time coordinate $t$ instead of $x_4$, and studied integral equations and the invariant transformation groups. He considered specifically transformations that leave the invariance of the differential (form) equation:

$$(dx)^2 + (dy)^2 + (dz)^2 - (dt)^2 = 0, \tag{5}$$

and include conformal transformations in addition to Lorentz transformations, therefore he went one step forward toward general coordinate invariance. (Without a generalization of Eq. (5), Bateman was not able to go one more step to general coordinate invariance. See also the third paragraph after Eq. (13) for a quote from Murnaghan's statement on Bateman's paper.)

With indefinite metric, one has to distinguish covariant and contravariant tensors and indices. Aware of this, one can readily put Maxwell equations into covariant form without using imaginary time. Following Minkowski[3] but use real time coordinate, in terms of Minkowski 4-dim field strength $F_{kl}$ (**E**, **B**) and 4-dim excitation (density) $H^{ij}$ (**D**, **H**):

$$F_{kl} = \begin{pmatrix} 0 & E_1 & E_2 & E_3 \\ -E_1 & 0 & -B_3 & B_2 \\ -E_2 & B_3 & 0 & -B_1 \\ -E_3 & -B_2 & B_1 & 0 \end{pmatrix}, \tag{6a}$$

$$H^{ij} = \begin{pmatrix} 0 & -D_1 & -D_2 & -D_3 \\ D_1 & 0 & -H_3 & H_2 \\ D_2 & H_3 & 0 & -H_1 \\ D_3 & -H_2 & H_1 & 0 \end{pmatrix}, \tag{6b}$$

Maxwell equations can be expressed in Minkowski form as

$$H^{ij}{}_{,j} = -4\pi J^i, \tag{7a}$$
$$e^{ijkl} F_{jk,l} = 0, \tag{7b}$$

where $J^k$ is the charge 4-current density ($\rho$, **J**) and $e^{ijkl}$ the completely anti-symmetric tensor density (Levi-Civita symbol) with $e^{0123} = 1$ (See, e. g., Hehl and Obukhov [7]).

The constitutive relation (2) between the excitation and the field can now be expressed in the form:

$$H^{ij} = \chi^{ij}(F_{kl}). \tag{8}$$

Here is $\chi^{ij}(F_{kl})$ is a functional with 6 independent degrees of freedom. For medium with a linear local response or in the linear local approximation, (8) reduced to

$$H^{ij} = \chi^{ijkl} F_{kl}, \tag{9}$$

with $\chi^{ijkl}$ the (linear) constitutive tensor density [8-11]. For isotropic dielectric and isotropic permeable medium, the constitutive tensor density has 2 degrees of freedom; for anisotropic dielectric and anisotropic permeable medium, the constitutive tensor density has 12 degrees of freedom; for general linear local medium (with magnetoelectric response), the constitutive tensor has 21 degrees of freedom (with $\chi^{ijkl} = \chi^{klij}$).

Introducing the metric $g_{ij}$ as gravitational potential in 1913 [12] and versed in general (coordinate-)covariant formalism in 1914 [13], Einstein put the Maxwell equations in general covariant form ($\mathcal{F}^{ij} = H^{ij}$ in our notation) [13]:

$$\mathcal{F}^{ij},_j = -4\pi J^i, \tag{10a}$$
$$F_{ij},_k + F_{jk},_i + F_{ki},_j = 0. \tag{10b}$$

Shortly after Einstein completed general relativity, Einstein noticed that the Maxwell equations can be formulated in a form independent of the metric gravitational potential in 1916 [14]. Einstein introduced the covariant $V$-six-vector Equations (10a) and (10b) which are independent of metric gravitational potential. Only the constitutive tensor density $\chi^{ijkl}$ is dependent on the metric gravitational potential $g^{ik}$ ($g = \det(g_{ij})$ and $g_{ij}$ the inverse of $g^{ij}$):

$$\mathcal{F}^{ij} = (-g)^{1/2} g^{ik} g^{jl} F_{kl}. \tag{11}$$

Noticing Einstein's $\mathcal{F}^{ij}$ is our $H^{ij}$ and putting (11) in the form of (9), we have

$$\chi^{ijkl} = (-g)^{1/2}[(1/2)g^{ik} g^{jl} - (1/2)g^{il} g^{kj}]. \tag{12}$$

In local inertial frame the metric-induced constitutive tensor (12) is reduced to special-relativitivistic Minkowski form ($\eta^{ik}$, Minkowski metric with signature $-2$):

$$\chi^{ijkl} = (-g)^{1/2}[(1/2)\eta^{ik} \eta^{jl} - (1/2)\eta^{il} \eta^{kj}] + O(x^i x^j). \tag{13}$$

Weyl[15] in his book on SPACE-TIME-MATTER used $F_{ik}$ and $\mathbf{F}^{ik}$ in writing Maxwell equations in general relativity in general covariant form. His $F_{ik}$ is just the electromagnetic field strength; his $\mathbf{F}^{ik}$ is related to $F_{ik}$ by equation (11) with $\mathbf{F}^{ij}$ replacing $\mathcal{F}^{ij}$. $F_{ik}$ is equal to $(\partial \phi_i/\partial x^k - \partial \phi_k/\partial x^i)$ with $\phi_i$'s (the electromagnetic 4-potential) the coefiecients of an invariant linear differential form (i.e., 1-form) $\phi_i dx^i$. Weyl went further in the penultimate section of his book to interpret the 1-form $\phi_i dx^i$ as scale (gauge) change form which is later tied to the phase of the four components of the wave field $\psi$ of electron.

There are two ways to show that Maxwell equations have general coordinate invariance: through differential calculus or through integral calculus. Mathmatically Maxwell equations depend only on the differential structure of spacetime, i.e. its differential topology. The above approach is basically through differential calculus.

Murnaghan in his article[16] in 1921 started with an elementary proof of a generalized Stokes' theorem for hyper-space, and stated Maxwell equations as integrals over hypersurfaces free of metric. While noting "Most of the results stated in this note are familiar to several writers on Mathematical Physics and the reader is referred to a paper by Bateman[5] for further developments" (footnote 1 on page 74 of [16]), he "hoped that the elementary presentation here given will enable physicists to become more easily familiar with the important results of this and similar papers". In this paper, he define the terms line integral, surface integral and hypersurface integral, not using any metrical properties of space; he also gave a simple derivation of the electromagnetic potential in hyperspace (spacetime). He explicitly and clearly showed that the Maxwell equations is metric free as he stated at the end of the first paragraph of his paper (page 74 of [16]): "However we shall have to speak of four dimensional space and as it is desirable to allow gravitational, i.e., non-euclidean spaces it may be permitted to recall the mathematical definition of space to show that the results obtained in no way depend on the metrical properties of the space.[1]"

This was followed by further stimulating and clarifying works of Kottler [17] in 1922 and Cartan [18] in 1923. Kottler in section 3 of his paper on "Maxwell's equation and metrics" established the independence of the prototype form of the Maxwell equations of any metric; the electromagnetic field (*E*) (N.B. in our nomenclature the 4-dim excitation) and the magneto-electric field (*F*) (N.B. in our nomenclature the 4-dim field strength) were regarded as previous-given quantities in an experiment. Kottler further wrote in section 6: "This led to the necessity of introducing constitutive relations between *E* and *F*, which establish interaction between electromagnetic and magneto-electric phenomena in space and time. These constitutive relations give rise to the appearance of a metric in Maxwell's equations, as their new form shows, e.g., (I *d*) and (II *d*) in matter-free space without gravity, according to Minkowski (with gravity, (I *c*) and (II *c*), according to Einstein). In sec. 5, it was then shown that for the Minkowskian (Einsteinian, resp.) vacuum the given of constitutive relations implies the given of the laws of radiation of the field in space and time, and conversely, the given of the latter implies the given of the former." Kottler's treatment was mainly using differencial calculus. Cartan put Maxwell equationw in forms and make the metric independence and the coordinate invariance transparent. For elaboration on Kottler's path and form approach, please see the review [1] in this issue.

In macroscopic medium, the constitutive tensor gives the medium-coupling to electromagnetism; it depends on the (thermodynamic) state of the medium and in turn depends on temperature, pressure etc. In gravity, the constitutive tensor gives the gravity-coupling to electromagnetism; it depends on the gravitational field(s) and in turn depends on the matter distribution and its state.

In gravity, a fundamental issue is how to arrive at the metric from the constitutive tensor in theory and through experiments/observations. That is, how to "derive" the metric theoretically, and how to build the metric empirically and test the EEP thoroughly. Are there other degrees of freedom to be explored?

Since ordinary energy is very small compared to Planck energy or the energy of Higgs partile and intermediate bosons, we can assume that the gravitational (or spacetime) constitutive tensor is a linear and local function of gravitational field(s), i.e. (9) holds in this situation. Since the second half of 1970's, we have started to use the following the Lagrangian density $L$ (= $L_I^{(EM)}$ + $L_I^{(EM-P)}$) with the electromagnetic field Lagrangian $L_I^{(EM)}$ and the field-current interaction Lagrangian $L_I^{(EM-P)}$ given by

$$L_I^{(EM)} = -(1/(16\pi))H^{ij} F_{ij} = -(1/(16\pi))\chi^{ijkl} F_{ij} F_{kl}, \quad (14a)$$
$$L_I^{(EM-P)} = -A_k J^k, \quad (14b)$$

for studying this issue [19-21]. Here $\chi^{ijkl} = -\chi^{jikl} = \chi^{klij}$ is a tensor density of the gravitational fields or matter fields to be investigated, $F_{ij} \equiv A_{j,i} - A_{i,j}$ the electromagnetic field strength tensor with $A_i$ the electromagnetic 4-potential and comma denoting partial derivation, and $J^k$ the charge 4-current density. The Maxwell equations (10a, b) or (1a-d) can be derived from this Lagrangian with the relation (12) and (6a,b). Using this $\chi$-framework, we have demonstrated the construction of the light cone core metric from the experiments and observations as in Table 1 [22, 23] with update of CPR fluctuation constraint to $<(\alpha − <\alpha>)^2>^{1/2} < 0.02$ from [24]. In the following sections, we summarize in more detail the construction of Table 1. Since the core theoretical constraint

$$\chi^{ijkl} = \tfrac{1}{2} (-h)^{1/2}[h^{ik} h^{jl} - h^{il} h^{kj}]\psi + \varphi e^{ijkl}, \quad (15)$$

is derived from the WEP I (Weak Equivalence Principle I), i.e. nonbirefringence, for photons/wave packets of light, we start with a discussion of WEP's in Setion 2. In (15), $h$ = det ($h_{ij}$) and $h_{ij}$ the inverse of $h^{ij}$ is a metric defined in terms of $\chi^{ijkl}$ up to a conformal factor. In section 3 after we briefly summarize the structure of premetric electrodynamics, we give a fairly detailed account of the construction of the core metric constraints from the WEP I for photons together with its empirical support using cosmic observations in this framework. In section 4, after presenting a brief history on Abelian (EM) axions and meaning of dilaton, we discuss wave propagation and dispersion relations in axion field and dilation field. Section 5 presents the Cosmic Polarization rotation (CPR) due to pseudoscalar-photon coupling, the amplification/attenuation effect due to dilaton coupling and their empirical constraints. Section 6 addresses the issue of empirical foundation of the closure relation. Section 7 summarizes the empirical constraints on skewons and discusses the special case of spacetime/medium with constitutive tensor induced by asymmetric metric. In Section 8, we give a "derivation" of Minkowski-Lorentz metric and elaborate further discussions.

Table 1. Constraints on the spacetime constitutive tensor $\chi^{ijkl}$ and construction of the spacetime structure (metric + axion field $\varphi$ + dilaton field $\psi$) from experiments/observations in skewonless case ($U$: Newtonian gravitational potential). $g_{ij}$ is the particle metric. [22-24]

| Experiment | Constraints | Accuracy |
|---|---|---|
| Pulsar Signal Propagation | | $10^{-16}$ |
| Radio Galaxy Observation | $\chi^{ijkl} \to \frac{1}{2} (-h)^{1/2}[h^{ik} h^{jl} - h^{il} h^{kj}]\psi + \varphi e^{ijkl}$ | $10^{-32}$ |
| Gamma Ray Burst (GRB) | | $10^{-38}$ |
| CMB Spectrum Measurement | $\psi \to 1$ | $8 \times 10^{-4}$ |
| Cosmic Polarization Rotation Experiment | $\varphi - \varphi_0 (\equiv \alpha) \to 0$ | $|<\alpha>| < 0.02$, $<(\alpha-<\alpha>)^2>^{1/2} < 0.02$ |
| Eötvös-Dicke-Braginsky Experiments | $\psi \to 1$ | $10^{-10}\, U$ |
| | $h_{00} \to g_{00}$ | $10^{-6}\, U$ |
| Vessot-Levine Redshift Experiment | $h_{00} \to g_{00}$ | $1.4 \times 10^{-4}\, \Delta U$ |
| Hughes-Drever-type Experiments | $h_{\mu\nu} \to g_{\mu\nu}$ | $10^{-24}$ |
| | $h_{0\mu} \to g_{0\nu}$ | $10^{-19}$ -$10^{-20}$ |
| | $h_{00} \to g_{00}$ | $10^{-16}$ |

## 2. Equivalence Principles for Photons/Wave Packets of Light[22,23]

(i) *WEP I for photons* (wave packets of light):

In analogue to the Galileo equivalence principle for test bodies, the WEP I for photons states that the spacetime trajectory of light in a gravitational field depends only on its initial position and direction of propagation, does not depend on its frequency (energy) and polarization. *This is equivalent to nonbirefringence of light propagation.*

(ii) *WEP II for photons* (wave packets of light):

The trajectory of light in a gravitational field depends only on its initial position and direction of propagation, not dependent on its frequency (energy) and polarization; the polarization state of the light does not change, e.g. no polarization rotation for linear polarized light; and no amplification/attenuation of light.

*Polarization rotation is relative to the bundle of light trajectories. Amplification/attenuation is relative to the counting of photons per moving volume of light bundles.*

N.B. We consider the propagation (or trajectory) in eikonal approximation, i.e. in geometrical optics approximation. The wavelength must be small (just like length scale of a test body) than the inhomogeneity scale of the gravitational field.

## 3. Gravitational Coupling to Electromagnetism and the Structure of Spacetime

In the early 1970's, a focus in Kip Thorne's group of theoretical astrophysics at Caltech is the study of various theories of gravity and the theoretical relationship among equivalence principles which I was interested. An issue is whether Schiff's conjecture[25] that Galileo Equivalence Principle implies Einstein Equivalence Principle is valid. After I moved to Montana, I spent some time in finding a counterexample and indeed found one.[26] The counterexample is an electromagnetic system with Lagrangian density

$$L = L_{I,GR}^{(EM)} + L_I^{(EM-P)} + L_I^{(NM)} + L_I^{(P)}, \tag{16}$$

where the electromagnetic field Lagrangian $L_{I,GR}^{(EM)}$ is given by (14a) with $\chi^{ijkl} = ½ (−g)^{1/2}[g^{ik} g^{jl} − g^{il} g^{kj}]$, the field-current interaction Lagrangian $L_I^{(EM-P)}$ is given by (14b), and the particle Lagangian $L_I^{(P)}$ is given by $− \Sigma_I m_I (ds_I/dt) \delta(\boldsymbol{x}−\boldsymbol{x}_I)$ with $m_I$ the mass of the $I$th particle, $e_I$ is the charge of the $I$th particle, $s_I$ its 4-line element from the metric $g_{ij}$, $\boldsymbol{x}_I$ its position 3-vector, $\boldsymbol{x}$ the space coordinates, $t$ the time coordinate, and the nonmetric part $L_I^{(NM)}$ is:[26-28]

$$L_I^{(NM)} = \eta e^{ijkl} A_i A_{j,k} \phi_{,l} \equiv −(1/4)\eta e^{ijkl} F_{ij} F_{kl} \phi \text{ (mod. div.)}. \tag{17}$$

The 4-current densiy $J^k$ of this system is given by

$$J^k = \Sigma_I e_I (dx_I^k/dt) \delta(\boldsymbol{x}−\boldsymbol{x}_I). \tag{18}$$

The total Lagrangian $L$ of this system written explicitly is

$$\begin{aligned}L &= L_{I,GR}^{(EM)} + L_I^{(EM-P)} + L_I^{(NM)} + L_I^{(P)} \\ &= − (1/(16\pi)) (−g)^{1/2} F^{ij}F_{kl} − A_k J^k − (1/4)\eta e^{ijkl} F_{ij} F_{kl} \phi − \Sigma_I m_I (ds_I/dt) \delta(\boldsymbol{x}−\boldsymbol{x}_I).\end{aligned} \tag{19}$$

We note that except for the last term $L_I^{(P)}$ this Lagrangian density is metric free (the $(−g)^{1/2}$ in the first term times $d^4x$ is the volume element. The $\varphi$ in the nonmetric Lagrangian (17) is sometimes called Abelian axion or EM axion (in [7], it is called Abelian axion), and the Lagrangian density $L_I^{(NM)}$ called the pseudoscalar-photon or axion-photon interaction term.

After I had written this nonmetric paper and distribute it out as a preprint, I immediately looked into the issue under what conditions Schiff's conjecture would be right. Aware of the constitutive tensor formalism in electrodynamics of continuous media, I thought it would be good to treat the constitutive tensor density as gravitational field and to find the conditions on it for Galileo equivalence principle to be valid. I worked it out and found that the nonmetric theory counterexample (19) [which can be included in the $L_I^{(EM)}$ (14a) + $L_I^{(EM-P)}$ (14b) + $L_I^{(P)}$ framework ($\chi$-$g$ framework) with $\chi^{ijkl} = ½ (−g)^{1/2}(g^{ik} g^{jl} − g^{il} g^{kj}) − (1/4) \eta e^{ijkl} F_{ij} F_{kl} \varphi]$ is the only counterexample in this framework.[27,28] $\phi$ is related to $\varphi$ of (15) by a constant multiple, i.e. $\varphi = −(1/4) \eta \phi$. However, to reach the conclusion from experiments, one needed also do Galileo equivalence principle experiments on electromagnetically polarized bodies, but this was lacking at the time. I proposed to do experiments on polarized bodies and searched for other empirical evidences.[28] I then turned into weak equivalence for photons, i.e. nonbirefringce and looked for data on pulsar pulse propagation. This way I started to use the $\chi$-framework (14a,b) for theoretical and phenomenological investigations since 1976.[19-21]

In subsection 3.1, we review wave propagation and the dispersion relations in the $\chi$-framework (14a,b). In subsection 3.2, we impose WEP I for photons (i.e. nonbirefringence condition) and derive the core result in this section. In subsection 3.3, we summarize the empirical constraints on the nonbirefringence.

### 3.1. Wave propagation and the dispersion relations

In (14a), the field intensity is defined through a 4-potential $A_i$ such that

$$F_{ij} = A_{j,i} - A_{i,j}, \tag{20}$$

with a gauge transformation freedom of adding an arbitrary gradient of a scalar function to $A_i$. The associated Maxwell equation in vacuum is

$$(\chi^{ijkl} A_{k,l})_{,j} = 0. \tag{21}$$

Using the derivation rule, we have

$$\chi^{ijkl} A_{k,lj} + \chi^{ijkl}{}_{,j} A_{k,l} = 0. \tag{22}$$

(i) For slowly varying, nearly homogeneous field/medium, and/or (ii) in the eikonal approximation with typical wavelength much smaller than the gradient scale and time-variation scale of the field/medium, the second term in (22) can be neglected compared to the first term, and we have

$$\chi^{ijkl} A_{k,lj} = 0. \tag{23}$$

This approximation is the lowest eikonal approximation, usually also called the eikonal approximation. In this approximation, the dispersion relation is given by the generalized covariant quartic Fresnel equation (see, e.g. [7], Eqs. (50)-(51) and section 7). It is well-known that the Abelian axion does not contribute to this dispersion relation [7, 19-22, 29-32] as we will see in the following. In this subsection, we use this lowest eikonal approximation and follow Refs. [19-21, 33] to derive dispersion relation in the general linear local constitutive framework. In the section 4, we keep the second term of (22) and follow Ref. [26, 34] to find out dispersion relations for the case that the dilaton gradient and the axion gradient cannot be neglected.

In the weak field or dilute medium, we assume

$$\chi^{ijkl} = \chi^{(0)ijkl} + \chi^{(1)ijkl} + O(2), \tag{24}$$

where O(2) means second order in $\chi^{(1)}$. Since the violation from the Einstein Equivalence Principle would be small and/or if the medium is dilute, in the following we assume that

$$\chi^{(0)ijkl} = (1/2) g^{ik} g^{jl} - (1/2) g^{il} g^{kj}, \tag{25}$$

and $\chi^{(1)ijkl}$ is small compared with $\chi^{(0)ijkl}$. We can then find a locally inertial frame such that $g^{ij}$ becomes the Minkowski metric $\eta^{ij}$ good to the derivative of the metric. To look for wave solutions, we use eikonal approximation and choose z-axis in the wave propagation direction so that the solution takes the following form:

$$A = (A_0, A_1, A_2, A_3)\, e^{ikz-i\omega t}. \tag{26}$$

We expand the solution as

$$A_i = [A^{(0)}{}_i + A^{(1)}{}_i + O(2)]\, e^{ikz-i\omega t}. \tag{27}$$

Imposing radiation gauge condition in the zeroth order in the weak field/dilute medium/weak EEP violation approximation, we find the zeroth order solution of (27) and the zeroth order dispersion relation satisfying the zeroth order equation $\chi^{(0)ijkl} A^{(0)}{}_{k,lj} = 0$ as follow:

$$A^{(0)} = (0, A^{(0)}{}_1, A^{(0)}{}_2, 0),\ \omega = k + O(1). \tag{28}$$

Substituting (24) and (27) into equation (23), we have

$$\chi^{(1)ijkl} A^{(0)}{}_{k,lj} + \chi^{(0)ijkl} A^{(1)}{}_{k,lj} = 0 + O(2). \tag{29}$$

The $i = 0$ and $i = 3$ components of (29) both give

$$A^{(1)}{}_0 + A^{(1)}{}_3 = 2\, (\chi^{(1)3013} - \chi^{(1)3010})\, A^{(0)}{}_1 + 2\, (\chi^{(1)3023} - \chi^{(1)3020})\, A^{(0)}{}_2 + O(2). \tag{30}$$

Since this equation does not contain $\omega$ and $k$, it does not contribute to the determination of the dispersion relation. A gauge condition in the O(1) order fixes the values of $A^{(1)}{}_0$ and $A^{(1)}{}_3$.

The $i = 1$ and $i = 2$ components of (29) are

$$(1/2)(\omega^2 - k^2)\, A^{(0)}{}_1 + \chi^{(0)1jkl} A^{(1)}{}_{k,lj} + \chi^{(0)1jkl} A^{(0)}{}_{k,lj} = 0 + O(2), \tag{31a}$$
$$(-1/2)(\omega^2 - k^2)\, A^{(0)}{}_2 + \chi^{(0)2jkl} A^{(1)}{}_{k,lj} + \chi^{(1)2jkl} A^{(0)}{}_{k,lj} = 0 + O(2). \tag{31b}$$

These two equations determine the dispersion relation and can be rewritten as

$$[(1/2)(\omega^2 - k^2) - k^2 A_{(1)}]\, A^{(0)}{}_1 - k^2 B_{(1)}\, A^{(0)}{}_2 = O(2), \tag{32a}$$
$$- k^2 B_{(2)}\, A^{(0)}{}_1 + [(1/2)(\omega^2 - k^2) - k^2 A_{(2)}]\, A^{(0)}{}_2 = O(2), \tag{32b}$$

where

$$A_{(1)} \equiv \chi^{(1)1010} - (\chi^{(1)1013} + \chi^{(1)1310}) + \chi^{(1)1313}, \tag{33a}$$
$$A_{(2)} \equiv \chi^{(1)2020} - (\chi^{(1)2023} + \chi^{(1)2320}) + \chi^{(1)2323}, \tag{33b}$$
$$B_{(1)} \equiv \chi^{(1)1020} - (\chi^{(1)1023} + \chi^{(1)1320}) + \chi^{(1)1323} \equiv B, \tag{33c}$$
$$B_{(2)} \equiv \chi^{(1)2010} - (\chi^{(1)2013} + \chi^{(1)2310}) + \chi^{(1)2313}. \tag{33d}$$

Since $\chi^{ijkl} = \chi^{klij}$, $B_{(1)} = B_{(2)} \equiv B$. We note that in $A_{(1)}$, $A_{(2)}$ *and B the axion part of $\chi$ drops* *out and, hence does not contribute to the dispersion relation in the lowest eikonal approximation.* For equations (32a,b) to have nontrivial solutions of $(A_1^{(0)}, A_2^{(0)})$, we must have the following determinant vanish to first order:

$$\det \begin{bmatrix} (1/2)(\omega^2 - k^2) - k^2 A_{(1)}] & -k^2 B \\ -k^2 B & (1/2)(\omega^2 - k^2) - k^2 A_{(2)} \end{bmatrix}$$
$$= (1/4)(\omega^2 - k^2)^2 - (1/2)(\omega^2 - k^2) k^2 (A_{(1)} + A_{(2)}) + k^4 (A_{(1)} A_{(2)} - B^2) = 0 + O(2). \quad (34)$$

The solution of this quadratic equation in $\omega^2$, i.e., the dispersion relation is

$$\omega^2 = k^2 [1 + (A_{(1)} + A_{(2)}) \pm ((A_{(1)} - A_{(2)})^2 + 4B^2)^{1/2}] + O(2), \quad (35)$$

or

$$\omega = k [1 + 1/2 (A_{(1)} + A_{(2)}) \pm 1/2 ((A_{(1)} - A_{(2)})^2 + 4 B^2)^{1/2}] + O(2). \quad (36)$$

From (39) the group velocity is

$$v_g = \partial \omega / \partial k = 1 + 1/2 (A_{(1)} + A_{(2)}) \pm 1/2 ((A_{(1)} - A_{(2)})^2 + 4 B^2)^{1/2} + O(2). \quad (37)$$

The quantity under the square root sign is

$$\xi \equiv (A_{(1)} - A_{(2)})^2 + 4 B^2. \quad (38)$$

For nonbirefringence, if and only if

$$A_{(1)} = A_{(2)}, B = 0. \quad (39)$$

### 3.1.1. The condition of vanishing B for all directions of wave propagation

From the definition (33c), the condition of vanishing of $B$ for wave propagation in the $z$-axis direction is

$$B = B_{(1)} = \chi^{(1)1020} + \chi^{(1)1323} - \chi^{(1)1023} - \chi^{(1)1320} = 0. \quad (40)$$

To look for conditions derivable in combination with those from other directions, we do active Lorentz transformations (rotations/boosts). Active rotation $R_\theta$ in the $y$-$z$ plane with angle $\theta$ is

$$\underline{t} = R_\theta t, \; \underline{x} = R_\theta x, \; \underline{y} = R_\theta y = y \cos\theta + z \sin\theta, \; \underline{z} = R_\theta z = -y \sin\theta + z \cos\theta. \quad (41)$$

Applying active rotation $R_\theta$ (41) to (40), we have

$$0 = \chi^{(1)1020} + \chi^{(1)1323} - \chi^{(1)1023} - \chi^{(1)1320}$$
$$= \chi^{(1)1020} + \chi^{(1)1323} - \chi^{(1)1023} - \chi^{(1)1320} + \theta\,(\chi^{(1)1030} + \chi^{(1)1220} - \chi^{(1)1223} - \chi^{(1)1330}) + O(\theta^2), \quad (42)$$

for small value of $\theta$. From (40) and (41), we have

$$\chi^{(1)1030} + \chi^{(1)1220} - \chi^{(1)1223} - \chi^{(1)1330} = 0. \quad (43)$$

Following the same procedure, we apply repeatedly active rotation $R_\theta$ to (43) and the resulting equations together with their linear combinations. After performing cyclic permutation 1→2→3→1 on the upper indices once and twice on some of the resulting equations, we have the following equations (for detailed derivation, see arXiv:1312.3056v1)

$$\chi^{(1)1220} = \chi^{(1)1330}; \; \chi^{(1)2330} = \chi^{(1)2110}; \; \chi^{(1)3110} = \chi^{(1)3220}; \; \chi^{(1)1020} = -\chi^{(1)1323}; \; \chi^{(1)2030} = -\chi^{(1)2131};$$
$$\chi^{(1)3010} = -\chi^{(1)3212}; \; \chi^{(1)1320} = -\chi^{(1)1230}; \; \chi^{(1)3210} = -\chi^{(1)3120}. \quad (44\text{a-h})$$

### 3.2. The nonbirefringence condition and the core metric theorem

With the condition $B = 0$ and $A_{(1)} = A_{(2)}$ for all directions of wave propagation, there is no birefringence for all directions of wave propagation. From subsection 3.1.1, we have equations (44a-h) holds from the validity of $B = 0$ for all directions of wave propagation. From $A_{(1)} = A_{(2)}$ and the definition (33a, b), we have

$$\chi^{(1)1010} - (\chi^{(1)1013} + \chi^{(1)1310}) + \chi^{(1)1313} = \chi^{(1)2020} - (\chi^{(1)2023} + \chi^{(1)2320}) + \chi^{(1)2323}. \quad (45)$$

From (44c) for the principal part, the terms in the parentheses on the two sides of the above equation cancel out and we have

$$\chi^{(1)1010} + \chi^{(1)1313} = \chi^{(1)2020} + \chi^{(1)2323}. \quad (46\text{a})$$

Applying active rotation $R_{\pi/2}$ around in the $y$-$z$ plane to (46a), we obtain

$$\chi^{(1)1010} + \chi^{(1)1212} = \chi^{(1)3030} + \chi^{(1)3232}. \quad (46\text{b})$$

With the conditions (39) in every directions, we derive (44a-h) and (46a,b). These 10 equations give ten conditions on 21 independent components of constitutive tensor density $\chi^{ijkl}$:

$$\chi^{(1)1220} = \chi^{(1)1330}; \quad (47\text{a})$$
$$\chi^{(1)2330} = \chi^{(1)2110}; \quad (47\text{b})$$
$$\chi^{(1)3110} = \chi^{(1)3220}; \quad (47\text{c})$$
$$\chi^{(1)1020} = -\chi^{(1)1323}; \quad (47\text{d})$$
$$\chi^{(1)2030} = -\chi^{(1)2131}; \quad (47\text{e})$$

$$\chi^{(1)3010} = -\chi^{(1)3212}; \tag{47f}$$

$$\chi^{(1)1320} = -\chi^{(1)1230}; \tag{47g}$$

$$\chi^{(1)3210} = -\chi^{(1)3120}; \tag{47h}$$

$$\chi^{(1)1010} + \chi^{(1)1313} = \chi^{(1)2020} + \chi^{(1)2323}; \tag{47i}$$

$$\chi^{(1)1010} + \chi^{(1)1212} = \chi^{(1)3030} + \chi^{(1)3232}. \tag{47j}$$

Define

$$h^{(1)10} \equiv h^{(1)01} \equiv -2\chi^{(1)1220};\ h^{(1)20} \equiv h^{(1)02} \equiv -2\chi^{(1)2330};\ h^{(1)30} \equiv h^{(1)03} \equiv -2\chi^{(1)3110};$$
$$h^{(1)12} \equiv h^{(1)21} \equiv -2\chi^{(1)1020};\ h^{(1)23} \equiv h^{(1)32} \equiv -2\chi^{(1)2030};\ h^{(1)31} \equiv h^{(1)13} \equiv -2\,^{(P)}\chi^{(1)3010};$$
$$h^{(1)11} \equiv 2\chi^{(1)2020} + 2\chi^{(1)2121} - h^{(1)00};\ h^{(1)22} \equiv 2\chi^{(1)3030} + 2\chi^{(1)3232} - h^{(1)00};$$
$$h^{(1)33} \equiv 2\chi^{(1)1010} + 2\chi^{(1)1313} - h^{(1)00}, \tag{48a}$$

$$\psi \equiv 1 + 2\chi^{(1)1212} + (1/2)\eta_{00}(h^{(1)00} - h^{(1)11} - h^{(1)22} - h^{(1)33}) - h^{(1)11} - h^{(1)22}, \tag{48b}$$

$$\varphi \equiv \chi^{(1)0123} \equiv \chi^{(1)[0123]}. \tag{48c}$$

Note that in these definitions, $h^{(1)00}$ is not defined and is free. Now it is straightforward to show that when (47a-j) are satisfied, then $\chi$ can be written to first-order in terms of the fields $h^{(1)ij}$, $\psi$ and $\varphi$ with $h^{ij} \equiv \eta^{ij} + h^{(1)ij}$ and $h \equiv \det(h_{ij})$ in the following form:

$$\chi^{ijkl} = \tfrac{1}{2}(-h)^{1/2}[h^{ik}h^{jl} - h^{il}h^{kj}]\psi + \varphi e^{ijkl}. \tag{49}$$

We are ready to derive the following theorem:

*Theorem*: For linear electrodynamics with Lagrangian (14a), the following three statements are equivalent to first order in the field:
(i) $A_{(1)} = A_{(2)}$ and $^{(P)}B = 0$ for all directions, i.e. nonbirefringence in electromagnetic wave propagation,
(ii) (47a-j) hold,
(iii) $\chi^{ijkl}$ can be expressed as (49) with (48a-c).

*Proof*: (i) → (ii) has been demonstrated in the derivation of (47a-j).
(ii) → (iii) has also been demonstrated in the derivation of (49) above.
(iii) → (i): (49) is a Lorentz tensor density equation. If it holds in one Lorentz frame, it holds in any other frame. From this we readily check that $A_{(1)} = A_{(2)}$ and $^{(P)}B = 0$ in any new frame with the wave propagation in the $z$-direction.

This theorem is a re-statement of the results of our work [19-21]. We note that previously we used the symbol $H^{ik}$ instead of $h^{ik}$, here because $H^{ik}$ is already used for electromagnetic excitation, we changed the notation.

We constructed the relation (49) in the weak-violation approximation of EEP in 1981 [19-21]; Haugan and Kauffmann [29] reconstructed the relation (49) in 1995. After the cornerstone work of Lämmerzahl and Hehl [30], Favaro and Bergamin [35] finally proved the relation (49) without assuming weak-field approximation (see also Dahl [36]; and [1] for a more detailed review). The strong field proof is very important, not just by itself but also because in ultrahigh precision empirical tests on nonbirefingence the strong field proof guarantees a good approximate metric to do the weak field analysis to give the precision nonbirefringenc constraints in each direction; in case there is a weak violation, our constructive approach will give the magnitude of violation in each direction. Briefly, the Lämmerzahl-Hehl-Favaro-Bergamin approach used the generalized covariant quartic Fresnel equation in the wave covector $q_i$ derived in the eikonal approximation in terms of electromagnetic field intensity (see, e.g. [7]):

$$G^{ijkl}(\chi)q_iq_jq_kq_l = 0, \qquad (50)$$

where $G^{ijkl}(\chi)$ (= $G^{(ijkl)}(\chi)$) is a completely symmetric fourth order Tamm-Rubilar (TR) tensor density of weight +1 defined by

$$G^{ijkl}(\chi) \equiv (1/4!)\,\underline{e}_{mnpq}\,\underline{e}_{rstu}\chi^{mnr(i}\chi^{j|ps|k}\chi^{l)qtu}. \qquad (51)$$

The parentheses in the upper indices mean symmetrization in $i$, $j$, $k$ and $l$. Solving (51) for dispersion relation and imposing the condition of nonbirefringce/WEP I for photons, the quartic equation degenerates into two identical quadratic factors. The quadratic factor then gives the propagating light cone.

### 3.3. Empirical constraints on the birefringence/violation of WEP I for photons

Polarization measurements of electromagnetic waves from pulsars [20, 21] and cosmologically distant astrophysical sources [29, 37] has yielded stringent constraints agreeing with (49) down to $10^{-16}$ and $10^{-32}$ respectively as shown in Table 1. However, the best constraints have been coming from gamma-ray burst observations.

*Constraints from gamma-ray burst observations* [22]: Recent polarization observations on gamma-ray bursts gives even better constraints on the dispersion relation and birefringence/violation of WEP I in cosmic propagation [38, 39]. The observation on the polarized gamma-ray burst GRB 061122 ($z = 1.33$) gives a lower limit on its polarization fraction of 60% at 68% confidence level (c.l.) and 33% at 90% c.l. in the 250-800 keV energy range [38]. The observation on the polarized gamma-ray burst GRB 140206A constrains the linear polarization level of the second peak of this GRB above 28 % at 90% c.l. in the 200-400 keV energy range [39]; the redshift of the source is measured from the GRB afterglow optical spectroscopy to be $z = 2.739$. GRBs polarization observations have been used to set constraints on various dispersion relations (See, e.g. [40, 41] and references therein). These two new GRB observations have larger and better redshift determinations than previous observations. We use them to give better constraints in our case. Since birefringence is proportional to the wave vector $k$ in our case, as gamma-ray of a particular frequency (energy) travels in the cosmic spacetime, the two linear polarization eigen-modes would pick up small phase differences. A linear polarization mode from distant source resolved into these two modes will become elliptical polarized during travel and lose part of the linear coherence. The way of gamma ray losing linear coherence depends on the frequency span. For a band of frequency, the extent of losing coherence depends on the distance of travel. The depolarization distance is of the order of frequency band span $\pi \Delta f$ times the integral $I = \int (1 + z(t))dt$ of the redshift factor $(1 + z(t))$ with respect to the time of travel. For GRB 140206A, this is about

$$\pi \Delta f\, I = \pi \Delta f \int (1 + z(t))dt \approx 1.5 \times 10^{20}\ \text{Hz} \times 0.6 \times 10^{18}\ \text{s} \approx 10^{38}. \qquad (52)$$

Since we do observe linear polarization in the 200-400 kHz frequency band of GRB 140206A with lower bound of 28 %, this gives a fractional constraint of about $10^{-38}$ on a combination of $\chi$'s. A similar constraint can be obtained for GRB 061122 (the band width times the redshift is about the same). A more detailed modeling may give better limits. The distribution of GRBs is basically isotropic. When this procedure is applied to an ensemble of polarized GRBs from various directions, the relation (49) would be verified to about $10^{-38}$.

Thus to high accuracy, photons are propagating in the metric field $h^{ik}$ with two additional (pseudo)scalar fields $\psi$ and $\varphi$. A change of $h^{ik}$ to $\lambda h^{ik}$ does not affect $\chi^{ijkl}$ in (49) -- this corresponds to the freedom of $h^{(1)00}$ in the definition (48a) of $h^{(1)ij}$. Thus we have constrained the general linear constitutive tensor of 21 degrees of freedom from the 10 constraints (47a-j) to eleven degrees of freedom in (49).

Therefore, we see that from the pulsar signal propagation, the polarization observations on radio galaxies and the gamma ray burst observations, the condition of nonbirefringence/WEP I for photons is verified empirically in spacetime propagarion with accuracies to $10^{-16}$, $10^{-32}$, and $10^{-38}$. The accuracies of these three observational constraints are summarized in Table I. The constitutive tensor can be constructed by the procedure in the proof of the theorem in this section to be in the core form (49) with accuracy to $10^{-38}$. Nonbirefringence (no splitting, no retardation) for electromagnetic wave propagation independent of polarization and frequency (energy) is the statement of Galileo Equivalence Principle for photons or WEP I for photons. Hence WEP I for photons is verified to this accuracy in the spacetime propagation.

In the following section, we assume (49) is valid and look into the influence of the axion field and the dilaton field of the constitutive tensor on the dispersion relation.

**4. Abelian Axions and Dilatons**

In 1950, Schrödinger [42] wrote a book on "Space-Time Structure" investigating the fundamental structure of spacetime and gravitation. The two basic principles of GR -- (i) Equivalence of all four-dimensional systems of coordinates obtained from any one of them by arbitrary (point-) transformation; (ii) the continuum has a metrical connexion impressed on it … -- were distinquished at the beginning (p. 2 of [42]). For (i), Schrödinger wrote "… there seems to be no reason to depart from it at the outset." For (ii), Schrödinger wrote "On the other hand, to adopt a metrical connexion straight away does not seem to be the simplest way of getting it eventually …" (p. 3 of [42]).

After obtaining the Maxwell equations (A') and (B') in the form Eq.'s (7a, b) or Eq.'s (10a, b) at the end of Part I of his book, Schrödinger wrote: "By (A') and (B') we have established Maxwell's fundamental equations invariantly in an arbitrary frame, using nothing but the means developed hitherto in these lectures; that is, for an unconnected space-time-manifold (neither affinity nor metric has been introduced). What we *cannot* establish in this manner is the relationship between the density (*H*, *D*) or $\mathfrak{f}^{ik}$ on the one side and the tensor (*B*, *E*) or $\phi_{ik}$ on theother side. (It is what in elementary theory is called the material equations.) For, the only relationship one could think of, to wit $\mathfrak{f}^{ik}$ = (1/2) $\varepsilon^{iklm}\varphi_{lm}$, makes the equations (A'), at least in the absence of current and charge ($\mathfrak{s}^k$ = 0), a consequence of (B') by identifying H with E and D with – B; which is entirely wrong and *could not be avoided by a different nomenclature*."

Here, Schrödinger considered the possibility of a constant Abelian axion constitutive tensor density and argued it was wrong. He considered it as the whole constitutive tensor density instead of a piece of the constitutive tensor density.

In the Lagrangian (14a), the constitutive tensor density $\chi^{ijkl}$ has 21 degrees of freedom (6 degrees of freedom for each pair (*i*, *j*) or (*k*, *l*), and hence $\chi^{ijkl}$ can be represented by 6 × 6 symmetric matrix which has 21 independent components). There is a special degree of freedom in this constitutive tensor which is totally antisymmetric in all indices. Post [43] made a theoretical argument in Chapter VI Section 2 of his book in 1962 that the alternating components in the constitutive tensor lead to an identically vanishing contribution in the Euler-Lagrangian derivative for matter by assuming that the gradient of the alternating components should be vanishing. Therefore by Post the number of independent elements of the Lagrangian-based constitutive tensor reduces to 20. This constraint is called Post constraint. There had been a controversy whether Post constraint should be valid in materials or not.

In 1957, Landau and Lifshitz [44] showed the possibility of the existence of a linear relationship between the electric field and magnetic field for materials of a certain type of magnetic crystal symmetry. In 1959, Dzyaloshinskii [45] showed that $Cr_2O_3$ has the right magnetic symmetry to have the magnetoelectric effect. In 1960-61, Astrov [46, 47] measured the magnetoelectric effect of $Cr_2O_3$ by applying an electric field. Wiegelmann, Jansen, Wyder, Rivera, and H. Schmid [48, 49] measured the magnetoelectric effect of $Cr_2O_3$ by applying a magnetic field. Based on the measurement of Astrov [47] and Wiegelmann *et al.* [48, 49], Hehl, Obukhov, Rivera and Schmid [50] showed in 2008 that the $Cr_2O_3$ crystal does have an axionic part of constitutive tensor and, hence settled the issue whether Post constraint should be valid in materials definitely: the Post constraint is not valid for materials. For some recent developments in the research on axionic materials, please see a summary in [1].

Dicke was not only a pioneer in reviving the experimental tests of gravitation, but also the senior proposer of Brans-Dicke scalar-tensor theory of gravity. In thought about possible role of scalar interaction, he wrote:[51]

"Consider first the scalar field. We can find at least two Lagragian densities (scalar density) which would represent a scalar field interacting with a vector field alone. They are:

$$\varphi^2 e^{ijkl}F_{ij}F_{kl} \text{ and } \varphi\varepsilon^{ijkl}F_{ij}A_k\varphi_{,l}. \quad (7)$$

As there is no self-interaction possible for the fields $\varphi$ and $A_i$, the Euler equations for each of the two fields must be linear in that field (generally inhomogeneneous but not necessarily in the other. Physically this means that for any vector field, assumed to be known and given, the equation of motion of $\varphi$ must be linear or else scalarons would not be linearly superposable, implying an interaction with each other. This linearity requirement implies that the Lagrangian density is quadratic in each of these fields leading to the unique form given by equation (7). We shall use the first form, keep in mind the other is possible."

"It must be said that this linearity requirement is not a strong condition as one could conceive of 3-body interactions in which a self-interaction of a field could take place only in the presence of a 2$^{nd}$ field, e.g., two scalarons could collide only in the presence of a vectoron. However, simplicity requires that we ignore such possibilities in the first approximation."

"As we have a self-interaction term for neither the scalar nor vector field, we must take the simplest form of action principle containing all the essential elements to be that with the first of equation (7) as the Lagrangian density. This leads to the Euler equations:

$$\varphi e^{ijkl}F_{ij}F_{kl} = 0 \quad (8)$$
$$\varphi\varphi_{,k}\varepsilon^{ijkl}F_{ij} = 0. \quad (9)$$

These are five equations for the five unknowns $\varphi$, $A_i$. However, because of the arbitrary coordinate system, four identities are satisfied by these equations, with only one equation remaining as nonredundant."

"A theory closer to the physical world as it actually exists can be constructed by considering the scalar field to interact with a tensor field only. …"

After these comments, Dicke went away to discuss scalar-tensor interactions and other interactions.

As we mentioned at the beginning of section 3, I arrived at the Abelian axion theory in 1973 by studying phenomenology among equivalence principles [26-28]. Weinberg [52] and Wilczek [53] proposed QCD axion theory following Peccei-Quinn [54]. Depending on models, QCD axions and/or string axions would or would not inudeuce Abelian axions. Thus both QCD axions and string axions are subject to the analysis and phenomenology following. For a review of QCD axions and string axions, please see [55, 56].

Dilaton may mean different things to different people. In Wiktionary, the meaning is: (i) (physics) A theoretical scalar field (analogous to the photon); (ii) (physics) A particle, associated with gravity, in string theory. It is related to scale change. Here, we mean specifically the scalar field $\psi$ in (49). For aspects of early history, see [57, 58]. Its effect on photon propagation will be derived in the following subsection.

*4.1. Wave propagation and the dispersion relation in axion field and dilaton field*

We first notice that in the lowest eikonal approximation, the dispersion relation (35) or (36) does not contain the axion piece and also does not contain the gradient of fields. Dilaton in (49) goes in this dispersion relation only as an overall scale factor and drops out too.

To derive the influence of the dilaton field and the axion field on the dispersion relation, one needs to keep the second term in equation (22). This has been done for the axion field in references [26, 31, 32, 59-61]. Here we follow the treatment in [34] to develop it for the joint dilaton field and axion field with the constitutive relation (49). Near the origin in a local inertial frame, the constitutive tensor density in dilaton field $\psi$ and axion field $\varphi$ [equation (49)] becomes

$$\chi^{ijkl}(x^m) = [(1/2)\, \eta^{ik} \eta^{jl} - (1/2)\, \eta^{il} \eta^{kj}]\, \psi(x^m) + \varphi(x^m)\, e^{ijkl} + O(\delta_{ij} x^i x^j), \qquad (53)$$

where $\eta^{ij}$ is the Minkowski metric with signature $-2$ and $\delta_{ij}$ the Kronecker delta. In the local inertial frame, we use the Minkowski metric and its inverse to raise and lower indices. Substituting (53) into the equation (22) and multiplying by 2, we have

$$\psi\, A^{i,j}{}_j + \psi\, A^{j,i}{}_j + \psi_{,j}\, A^{i,j} - \psi_{,j}\, A^{j,i} + 2\, \varphi_{,j}\, e^{ijkl}\, A_{k,l} = 0. \qquad (54)$$

We notice that (54) is both Lorentz covariant and gauge invariant.

We expand the dilaton field $\psi(x^m)$ and the axion field $\varphi(x^m)$ at the 4-point (event) $P$ with respect to the event (time and position) $P_0$ at the origin as follows:

$$\psi(x^m) = \psi(P_0) + \psi_{,i}(P_0)\, x^j + O(\delta_{ij} x^i x^j), \qquad (55a)$$
$$\varphi(x^m) = \varphi(P_0) + \varphi_{,i}(P_0)\, x^j + O(\delta_{ij} x^i x^j). \qquad (55b)$$

To look for wave solutions, we use eikonal approximation which does not neglect field gradient/medium inhomogeneity. Choose $z$-axis in the wave propagation direction so that the solution takes the following form:

$$A \equiv (A_0, A_1, A_2, A_3) = (\underline{A}_0, \underline{A}_1, \underline{A}_2, \underline{A}_3)\, e^{ikz-i\omega t} = \underline{A}_j\, e^{ikz-i\omega t}. \qquad (56)$$

Expand the solution as

$$A_i = A^{(0)}{}_i + A^{(1)}{}_i + O(2) = [\underline{A}^{(0)}{}_i + \underline{A}^{(1)}{}_i + O(2)]\, e^{ikz-i\omega t} = \underline{A}_i\, e^{ikz-i\omega t}. \qquad (57)$$

Now use eikonal approximation to obtain a local dispersion relation. In the eikonal approximation, we only keep terms linear in the derivative of the dilaton field and the axion field; we neglect terms containing the second-order derivatives of the dilaton field or the axion field, terms of $O(\delta_{ij} x^i x^j)$ and terms of mixed second order, e.g. terms of $O(A^{(1)}{}_i\, x^j)$ or $O(A^{(1)}{}_i\, \psi_{,j})$; we call all these terms $O(2)$.

Imposing radiation gauge condition in the zeroth order, finding the corresponding zeroth order solution and the dispersion relation, and iterating with suitable gauge condition in the first order, we obtain the following equations for $\underline{A}^{(0)}{}_1$ and $\underline{A}^{(0)}{}_2$:

$$(\omega^2 - k^2)\,\underline{A}^{(0)}{}_1 - i\,k\,\underline{A}^{(0)}{}_1\,\psi^{-1}(\psi_{,0} + \psi_{,3}) - 2\,i\,k\,\underline{A}^{(0)}{}_2\,\psi^{-1}(\varphi_{,0} + \varphi_{,3}) = 0 + O(2), \quad (58a)$$
$$(\omega^2 - k^2)\,\underline{A}^{(0)}{}_2 - i\,k\,\underline{A}^{(0)}{}_2\,\psi^{-1}(\psi_{,0} + \psi_{,3}) + 2\,i\,k\,\underline{A}^{(0)}{}_1\,\psi^{-1}(\varphi_{,0} + \varphi_{,3}) = 0 + O(2). \quad (58b)$$

For these two equations have nontrivial solutions, we must have the following determinant vanish:

$$\det \begin{bmatrix} (\omega^2 - k^2) - i\,k\,\psi^{-1}(\psi_{,0} + \psi_{,3}) & -2\,i\,k\,\psi^{-1}(\varphi_{,0} + \varphi_{,3}) \\ 2\,i\,k\,\psi^{-1}(\varphi_{,0} + \varphi_{,3}) & (\omega^2 - k^2) - i\,k\,\psi^{-1}(\psi_{,0} + \psi_{,3}) \end{bmatrix}$$
$$= [(\omega^2 - k^2) - i\,k\,\psi^{-1}(\psi_{,0} + \psi_{,3})]^2 - 4\,k^2\,\psi^{-2}(\varphi_{,0} + \varphi_{,3})^2 = 0 + O(2). \quad (59)$$

This is the dispersion relation in the axion field and the dilaton field. Its solution is

$$\omega = k - (i/2)\,\psi^{-1}(\psi_{,0} + \psi_{,3}) \pm \psi^{-1}(\varphi_{,0} + \varphi_{,3}) + O(2), \quad (60)$$

with the group velocity $v_g = \partial\omega/\partial k = 1$ independent of polarization. When the dispersion relation is satisfied, there are two independent solutions with the polarization eigenvectors $\underline{A}^{(0)}{}_i = (\underline{A}^{(0)}{}_0, \underline{A}^{(0)}{}_1, \underline{A}^{(0)}{}_2, \underline{A}^{(0)}{}_3)$:

$$\underline{A}^{(0)}{}_1 / \underline{A}^{(0)}{}_2 = [2\,i\,k\,\psi^{-1}(\varphi_{,0} + \varphi_{,3})] / [(\omega^2 - k^2) - i\,k\,\psi^{-1}(\psi_{,0} + \psi_{,3})]$$
$$= [2\,i\,k\,\psi^{-1}(\varphi_{,0} + \varphi_{,3})] / [\pm 2\,k\,\psi^{-1}(\varphi_{,0} + \varphi_{,3})] = \pm i; \quad (61a)$$
$$\underline{A}^{(0)}{}_0 = \underline{A}^{(0)}{}_3 = 0, \quad (61b)$$

for $\omega = k - (i/2)\,\psi^{-1}(\psi_{,0} + \psi_{,3}) \pm \psi^{-1}(\varphi_{,0} + \varphi_{,3}) + O(2)$ respectively. From (61a), the two polarization eigenstates are left circularly polarized state and right circularly polarized state in axion field. This agrees with the electromagnetic wave propagation in axion field as derived earlier [26, 31, 32, 59-61].

With the dispersion (60), the plane-wave solution (56) propagating in the $z$-direction is

$$A \equiv (A_0, A_1, A_2, A_3) = (0, \underline{A}^{(0)}{}_1, \underline{A}^{(0)}{}_2, 0)\,e^{ikz - i\omega t}$$
$$= (0, \underline{A}^{(0)}{}_1, \underline{A}^{(0)}{}_2, 0)\,\exp[ikz - ikt \pm (-i)\,\psi^{-1}(\varphi_{,0}\,t + \varphi_{,3}\,z) - (1/2)\,\psi^{-1}(\psi_{,0}\,t + \psi_{,3}\,z)], \quad (62)$$

with $\underline{A}^{(0)}{}_1 = \pm i\,\underline{A}^{(0)}{}_2$. The additional factor acquired in the propagation is $\exp[\pm(-i)\,\psi^{-1}(\varphi_{,0}\,t + \varphi_{,3}\,z)] \times \exp[-(1/2)\psi^{-1}(\psi_{,0}\,t + \psi_{,3}\,z)]$. The first part of this factor, i.e., the axion factor $\exp[\pm(-i)\,\psi^{-1}(\varphi_{,0}\,t + \varphi_{,3}\,z)]$ adds a phase in the propagation. The second part of this factor, i.e., the dilaton factor $\exp[-(1/2)\,\psi^{-1}(\psi_{,0}\,t + \psi_{,3}\,z)]$ amplifies or attenuates the wave according to whether $(\psi_{,0}\,t + \psi_{,3}\,z)$ is less than zero or greater than zero. For the right circularly polarized electromagnetic wave, the effect of the axion field in the propagation from a point $P_1 = \{x_{(1)}{}^i\} = \{x_{(1)}{}^0;\,x_{(1)}{}^\mu\} = \{x_{(1)}{}^0, x_{(1)}{}^1, x_{(1)}{}^2, x_{(1)}{}^3\}$ to another point $P_2 = \{x_{(2)}{}^i\} = \{x_{(2)}{}^0;\,x_{(2)}{}^\mu\} = \{x_{(2)}{}^0, x_{(2)}{}^1, x_{(2)}{}^2, x_{(2)}{}^3\}$ is to add a phase of $\alpha = \psi^{-1}\,[\varphi(P_2)$ $- \varphi(P_1)] \ (\approx \varphi(P_2) - \varphi(P_1)$ for $\psi \approx 1$) to the wave; for left circularly polarized light, the effect is to add an opposite phase [26, 31, 32, 59-61]. Linearly polarized electromagnetic wave is a superposition of circularly polarized waves. Its polarization vector will then rotate by an angle $\alpha$. The effect of the dilaton field is to amplify with a factor $\exp[-(1/2) \psi^{-1} (\psi_{,0} t + \psi_{,3} z)] = \exp[-(1/2) ((\ln \psi)_{,0} t + (\ln \psi)_{,3} z)] = (\psi(P_1)/\psi(P_2))^{1/2}$. The dilaton field contributes to the amplitude of the propagating wave is positive or negative depending on $\psi(P_1)/\psi(P_2) > 1$ or $\psi(P_1)/\psi(P_2) < 1$ respectively.

For plane wave propagating in direction $n^\mu = (n^1, n^2, n^3)$ with $(n^1)^2 + (n^2)^2 + (n^3)^2 = 1$, the solution is

$$A(n^\mu) \equiv (A_0, A_1, A_2, A_3) = (0, \underline{A}_1, \underline{A}_2, \underline{A}_3) \exp(-i\, kn^\mu x_\mu - i\omega t)$$
$$= (0, \underline{A}_1, \underline{A}_2, \underline{A}_3) \exp[-ikn^\mu x_\mu - ikt \pm (-i)\psi^{-1}(\varphi_{,0}t - n^\mu \varphi_{,\mu} n_\nu x^\nu) - (1/2)\, \psi^{-1}(\psi_{,0}t + n^\mu \psi_{,\mu} n_\nu x^\nu)], \quad (63)$$

where $\underline{A}_\mu = \underline{A}^{(0)}_\mu + n_\mu n^\nu \underline{A}^{(0)}_\nu$ with $\underline{A}^{(0)}_1 = \pm i\, \underline{A}^{(0)}_2$ and $\underline{A}^{(0)}_3 = 0$ $[n_\mu \equiv (-n^1, -n^2, -n^3)]$. There are polarization rotation for linearly polarized light due to axion field gradient, and amplification/attenuation due to dilaton field gradient.

The above analysis is local. In the global situation, choose local inertial frames along the wave trajectory and integrate along the trajectory. Since $\psi$ is a scalar, the integration gives $(\psi(P_1)/\psi(P_2))^{1/2}$ as the amplification factor for the propagation in the dilaton field. For small dilaton field variations, the amplification/attenuation factor is equal to $[1 - (1/2) (\Delta\psi/\psi)]$ to a very good approximation with $\Delta\psi \equiv \psi(P_2) - \psi(P_1)$. Since this factor does not depend on the wave number/frequency and polarization, it will not distort the source spectrum in propagation, but gives an overall amplification/attenuation factor to the spectrum. The axion field contributes to the phase factor and induces polarization rotation as in previous investigations [26, 31, 32, 59-61]. For $\psi \approx 1$ (constant), the induced polarization rotation agrees with previous results which were obtained without considering dilaton effect. If the dilaton field varies significantly, a $\psi$-weight needs to be included in the integration.

In the next section, we look into the empirical constraints on possible polarization rotation and amplification/attenuation.

## 5. Cosmic Amplification/Attenuation and Cosmic Polarization Rotation (CPR) in Dilaton Field and Axion Field and their Empirical Constraints

In this section we look into the observations/experiments to constrain the dilaton field contribution and the axion field contribution to spacetime constitutive tensor density following the exposition of Ref. [34]..

*No amplification/no attenuation constraint on the cosmic field*: From equation (62) and (63) in the last section, we have derived that the amplitude and phase factor of propagation in the cosmic dilaton and cosmic axion field is changed by $(\psi(P_1)/\psi(P_2))^{1/2} \times \exp[-ikn^\mu x_\mu - ikt \pm (-i)(\varphi(P_1) - \varphi(P_2))t]$. The effect of dilaton field is to give amplification $((\psi(P_1) - \psi(P_2) > 0)$ or attenuation $((\psi(P_1) - \psi(P_2) < 0)$ to the amplitude of the wave independent of frequency and polarization.

The spectrum of the cosmic microwave background (CMB) is well understood to be Planck blackbody spectrum. In the cosmic propagation, this spectrum would be amplified or attenuated by the factor $(\psi(P_1)/\psi(P_2))^{1/2}$. However, the CMB spectrum is measured to agree with the ideal Planck spectrum at temperature $2.7255 \pm 0.0006$ K [62] with a fractional accuracy of $2 \times 10^{-4}$. The spectrum is also red-shifted due to cosmological curvature (or expansion), but this does not change the blackbody character. The measured shape of the CMB spectra does not deviate from Planck spectrum within its experimental accuracy. In the dilaton field the relative increase in power is proportional to the amplitude increase squared, i.e., $\psi(P_1)/\psi(P_2)$. Since the total power of the blackbody radiation is proportional to the temperature to the fourth power $T^4$, the fractional change of the dilaton field since the last scattering surface of the CMB must be less than about $8 \times 10^{-4}$ and we have

$$|\Delta\psi|/\psi \leq 4\,(0.0006/2.7255) \approx 8 \times 10^{-4}. \qquad (64)$$

Direct fitting to the CMB data with the addition of the scale factor $\psi(P_1)/\psi(P_2)$ would give a more accurate value.

*Constraints on the cosmic polarization rotation (CPR) and the cosmic axion field*: From (63), for the right circularly polarized electromagnetic wave, the propagation from a point $P_1$ (4-point) to another point $P_2$ adds a phase of $\alpha = \varphi(P_2) - \varphi(P_1)$ to the wave; for left circularly polarized light, the added phase will be opposite in sign [26]. Linearly polarized electromagnetic wave is a superposition of circularly polarized waves. Its polarization vector will then rotate by an angle $\alpha$. In the global situation, it is the property of (pseudo)scalar field that when we integrate along light (wave) trajectory the total polarization rotation (relative to no $\varphi$-interaction) is again $\alpha = \Delta\varphi = \varphi(P_2) - \varphi(P_1)$ where $\varphi(P_1)$ and $\varphi(P_2)$ are the values of the scalar field at the beginning and end of the wave. This polarization rotation is called Cosmic Polarization Rotation (CPR). For vector field ($V_i$) coupling Lagrangian of the form $e^{ijkl}A_iA_{j,k}V_l$, Carroll, Field and Jackiw[63] also derived the CPR effect; locally it is proportional to $V_l dx^l$. If $V_l$ can be expressed as the gradient of a scalar, then CPR can be expressed as the scalar field difference between the beginning and the observation of the electromagnetic wave by simple integration along the photon 4-path. Alexander [64] and Caldwell [65] have accounted various recent theories/models which leads to observable CPRs. Gubitosi [66] addresses the issue of disentangling CPR/birefringence from standard physics in CMB measurements and distinguishing among production mechanisms,

The constraints listed on CPR for the axion field (or other field in various theories/models) are from the UV polarization observations of radio galaxies and the CMB polarization observations -- 0.02 for CPR mean value $|\langle\alpha\rangle|$ and 0.02 for the CPR fluctuations $\langle(\alpha - \langle\alpha\rangle)^2\rangle^{1/2}$. See di Serego Alighieri [67] for a review and summary. For most recent results, see [24, 68-70]. The most recent analysis of Planck CMB polarization data (30–353 GHz) gives $\langle\alpha\rangle = 0.35° \pm 0.05° \pm 0.28° = 0.0061 \pm 0.0009 \pm 0.0049$ [rad] with uniformity assessment of all-sky [69]. From the observations, the constraints on time variation is of the same order of magnitude as the space fluctuations. It may mean our universe is pretty in equilibrium as far as $\varphi$ is concerned. Therefore the measurement of its spatial fluctuations is even more important.

In our original pseudoscalar model [31, 32, 59], the natural coupling strength ϕ is of order 1. However, the isotropy of our observable universe to $10^{-5}$ may lead to a change of $\Delta\varphi$ over cosmological distance scale $10^{-5}$ smaller. Hence, observations to test and measure $\Delta\varphi$ to $10^{-6}$ are very significant. A positive result may indicate that our patch of inflationary universe has a 'spontaneous polarization' in the fundamental law of electromagnetic propagation influenced by neighboring patches and by a determination of this fundamental physical law we could 'observe' our neighboring patches.

For measurement reaching a polarization angle of $10^{-5}$-$10^{-6}$, calibration angle accuracy is very important. Galluzzi, M. Massardi *et al*. [71, 72] have looked into the properties of polarimetric multi-frequency radio sources related to telescope calibration issues. Kaufman *et al*. [73] looked into the potential of using the Crab Nebula as a high precision calibrator for CMB polarimeters. Johnson *et al*. [74, 75] proposed a CubeSat for calibrating ground-based and sub-orbital millimeter-wave polarimeters. de Bernardis and Masi [76] discussed possible improvement in the near future on temperature sensitivity and polarization calibration and conclude that: (i) the Large Scale Polarization Explorer (LSPE, http://planck.roma1.infn.it\lspe) on stratospheric balloon platform is expected to constrain the tensor-to-scalar reatio *r* for B-modes with an error $\sigma_r < 0.01$ and to have the final survey sensitivity for CPR ≤ 0.1° (1$\sigma$); (ii) Cosmic Origins Explorer (COrE, a proposal for ESA's M4 space mission) is expected to measure CPR very well, with a final uncertainty in the CPR ≤ 0.01° (1$\sigma$) (that is better than $2 \times 10^{-4}$[rad]), due to the accuracy of calibration and very high sensitivity.

*Additional constraints to have the unique physical metric*: From (64) the fractional change of dilaton $|\Delta\psi|/\psi$ is less than about $8 \times 10^{-4}$ since the time of the last scattering surface of the CMB. Eötvös-type experiments constrain the fractional variation of dilaton to $\sim 10^{-10}$ *U* where *U* is the dimensionless Newtonian potential in the experimental environment. Vessot-Levine redshift experiment and Hughes-Drever-type experiments give further constraints [32]. All these constraints are summarized in Table 1. This leads to unique physical metric to high precision for all degrees of freedom except the axion degree of freedom and cosmic dilaton degree of freedom which are only mildly constrained.

## 6. Empirical Foundation of the Closure Relation [22, 23]

In this section, we look into the empirical foundation of the closure relation for electrodynamics.

There are two equivalent definitions of constitutive tensor which are useful in various discussions (see, e. g., Hehl and Obukhov [7]). The first one is to take a dual on the first 2 indices of $\chi^{ijkl}$:

$$\kappa_{ij}{}^{kl} \equiv (1/2)\underline{e}_{ijmn} \chi^{mnkl}, \tag{65}$$

where $\underline{e}_{ijmn}$ is the completely antisymmetric tensor density of weight −1 with $\underline{e}_{0123} = 1$. Since $e_{ijmn}$ is a tensor density of weight −1 and $\chi^{mnkl}$ a tensor density of weight +1, $\kappa_{ij}{}^{kl}$ is a (twisted) tensor. From (65), we have

$$\chi^{mnkl} = (1/2)e^{ijmn}\kappa_{ij}{}^{kl}. \tag{66}$$

With this definition of constitutive tensor $\kappa_{ij}{}^{kl}$, the constitutive relation (9) becomes

$$*H_{ij} = \kappa_{ij}{}^{kl} F_{kl}, \tag{67}$$

where $*H_{ij}$ is the dual of $H^{ij}$, i.e.

$$*H_{ij} \equiv (1/2)\, \underline{e}_{ijmn} H^{mn}. \tag{68}$$

The second equivalent definition of the constitutive tensor is to use a $6 \times 6$ matrix representation $\kappa_I{}^J$. Since $\kappa_{ij}{}^{kl}$ is nonzero only when the antisymmetric pairs of indices ($ij$) and ($kl$) have values (01), (02), (03), (23), (31), (12), these index pairs can be enumerated by capital letters $I, J, \ldots$ from 1 to 6 to obtain $\kappa_I{}^J$ ($\equiv \kappa_{ij}{}^{kl}$). With the relabeling, $F_{ij} \rightarrow F_I$, $H^{ij} \rightarrow H^I$, $\underline{e}_{ijmn} \rightarrow \underline{e}_{IJ}$, $e^{ijmn} \rightarrow e^{IJ}$. We have $F_I = (\mathbf{E}, -\mathbf{B})$ and $(*H)_I = (-\mathbf{H}, -\mathbf{D})$. $\underline{e}_{IJ}$ and $e^{IJ}$ can be expressed in matrix form as

$$\underline{e}_{IJ} = e^{IJ} = \begin{bmatrix} 0 & \mathbf{I}_3 \\ \mathbf{I}_3 & 0 \end{bmatrix}, \tag{69}$$

where $\mathbf{I}_3$ is the $3 \times 3$ unit matrix. In terms of this definition, the constitutive relation (104) becomes

$$*H_I = 2\, \kappa_I{}^J F_J, \tag{70}$$

where $*H_I \equiv *H_{ij} = e_{IJ} H^J$. The axion part $^{(\mathrm{Ax})}\chi^{ijkl}$ ($= \varphi\, e^{ijkl}$) now corresponds to

$$^{(\mathrm{Ax})}\kappa_I{}^J = \varphi \begin{bmatrix} \mathbf{I}_3 & 0 \\ 0 & \mathbf{I}_3 \end{bmatrix} = \varphi\, \mathbf{I}_6, \tag{71}$$

where $\mathbf{I}_6$ is the $6 \times 6$ unit matrix. The principal part and the axion part of the constitutive tensor all satisfy the following equation (the skewonless condition):

$$e^{KJ}\kappa_J{}^I = e^{IJ}\, \kappa_J{}^K. \tag{72}$$

In terms of $\kappa_{ij}{}^{kl}$ and re-indexed $\kappa_I{}^J$, the constitutive tensor (15) or (49) is represented in the following forms:

$$\kappa_{ij}{}^{kl} = (1/2)\, \underline{e}_{ijmn}\, \chi^{mnkl} = (1/2)\, \underline{e}_{ijmn}\, (-h)^{1/2}\, h^{mk}\, h^{nl}\, \psi + \varphi\, \delta_{ij}{}^{kl}, \tag{73}$$
$$\kappa_I{}^J = (1/2)\, \underline{e}_{ijmn}\, (-h)^{1/2}\, h^{mk}\, h^{nl}\, \psi + \varphi\, \delta_I{}^J, \tag{74}$$

where $\delta_{ij}{}^{kl}$ is a generalized Kronecker delta defined as

$$\delta_{ij}{}^{kl} = \delta_i^k \delta_j^l - \delta_i^l \delta_j^k. \tag{75}$$

In the derivation, we have used the formula

$$\underline{e}_{ijmn} e^{mnkl} = 2 \delta_{ij}{}^{kl}. \tag{76}$$

Let us calculate $\kappa_{ij}{}^{kl}\kappa_{kl}{}^{pq}$ for the constitutive tensor (73):

$$\begin{aligned}\kappa_{ij}{}^{kl} \kappa_{kl}{}^{pq} &= [(1/2) \underline{e}_{ijmn} (-h)^{1/2} h^{mk} h^{nl} \psi + \varphi \, \delta_{ij}{}^{kl}] \, [(1/2) \underline{e}_{klrs} (-h)^{1/2} h^{rp} h^{sq} \psi + \varphi \, \delta_{kl}{}^{pq}] \\ &= -(1/2) \delta_{ij}{}^{pq}\psi^2 + 2 \delta_{ij}{}^{pq}\varphi^2 + 2 \underline{e}_{ijrs} (-h)^{1/2} h^{rp} h^{sq} \varphi \, \psi \\ &= -(1/2) \delta_{ij}{}^{pq}\psi^2 + 4 \varphi \, {}^{(P)}\kappa_{ij}{}^{pq} - 2 \delta_{ij}{}^{pq} \varphi^2, \end{aligned} \tag{77}$$

where ${}^{(P)}\kappa_{ij}{}^{pq}$ denotes the principal part of $\kappa_{ij}{}^{pq}$ (i.e., excluding the axion part) and where we have used (76) and the following relations

$$e_{klrs} h^{mk} h^{nl} h^{rp} h^{sq} = e^{mnpq} \det(h^{uv}), \tag{78}$$
$$\det(h^{uv}) = [\det(h_{uv})]^{-1} = h^{-1}, \tag{79}$$
$$\delta_{ij}{}^{kl} \delta_{kl}{}^{pq} = 2 \delta_{ij}{}^{pq}. \tag{80}$$

In terms of the six-dimensional index $I$, equation (77) becomes

$$\kappa_I{}^J \kappa_J{}^K = (1/2)\kappa_{ij}{}^{kl} \kappa_{kl}{}^{pq} = -(1/4)\psi^2\delta_I^K + 2{}^{(P)}\kappa_I{}^K \varphi - \delta_{ij}{}^{pq} \varphi^2 = -(1/4)\psi^2 \delta_I^K + 2{}^{(P)}\kappa_I{}^K \varphi - \delta_I^K \varphi^2. \tag{81}$$

Thus the matrix multiplication of $\kappa_I{}^J$ with itself is a linear combination of itself and the identity matrix, and generates a closed algebra of linear dimension 2. The algebraic relation (81) is a closure relation that generalizes the following closure relation in electrodynamics:

$$\kappa \, \kappa = (\kappa_I{}^J \kappa_J{}^K) = (1/6) \, \text{tr}(\kappa \, \kappa) \, \mathbf{I}_6. \tag{82}$$

In case $\varphi = 0$, the axion part ${}^{(Ax)}\kappa_I{}^J$ of the constitutive tensor vanishes and (81) reduces to the closure relation (82).

From the nonbirefringence condition (15) or (49), we derive the closure relation (81) in a number of algebraic steps which consist of order 100 individual operations of addition/subtraction or multiplication. Equation (15) or (49) is empirically verified to $10^{-38}$. Therefore equation (81) is empirically verified to $10^{-37}$ (precision $10^{-38}$ times $100^{1/2}$). Hence, when there are no axion and no dilaton, the closure relation (82) is empirically verified to $10^{-37}$. For dilaton is constrained to $8 \times 10^{-4}$, if one allow for dilaton, relation (82) is verified to $8 \times 10^{-4}$ since the last scattering surface of CMB; for axion is constrained to $10^{-2}$, if one allow for axion in addition, relation (82) is verified to $10^{-2}$ since the last scattering surface of CMB.

The closure relation (82) can also be called idempotent condition for it states that the multiplication of $\kappa$ by itself goes back essentially to itself. Toupin [77], Schonberg [78], and Jadczyk [79] in their theoretical approach started from this condition to obtain metric induced constitutive tensor with a dilaton degree of freedom. In this section, we have started with Galileo equivalence principle for photons, i.e. the nonbirefringence condition, to obtain the metric induced core metric form with a dilaton degree of freedom and an axion degree of freedom for the constitutive tensor and then the generalized closure relation (81). We have also shown that (81) is verified empirically to very high precision. Thus in the axionless case, the birefringence condition and idempotent condition are equivalent and both are verified empirically to high precision.

**7. Spacetime Constitutive Relation including Skewons** [22, 33]

If we allow medium with dissipation or amplification which does not need to have a Lagrangian (like absorptive medium or lasing medium), then the constitutive tensor $\chi^{ijkl}$ does not need to have the symmetry $\chi^{ijkl} = \chi^{klij}$. In this case the constitutive tensor has 36 (=6 × 6) degrees of freedom. Besides the the principal part (P) which has 20 degrees of freedom and the axion part (Ax) which has 1 degrees of freedom, the skewon part has 15 degrees of freedom. The principal part (P), the axion part (Ax) and the Hehl-Obukhov-Rubilar skewon part (Sk) (antisymmetric under exchange of pair indices (*ij*) and (*kl*).constitute the three irreducible parts under the group of general coordinate transformations [7]:

$$\chi^{ijkl} = {}^{(P)}\chi^{ijkl} + {}^{(Ax)}\chi^{ijkl} + {}^{(Sk)}\chi^{ijkl}, \quad (\chi^{ijkl} = -\chi^{jikl} = -\chi^{ijlk}) \qquad (83)$$

with

$$
\begin{align}
{}^{(P)}\chi^{ijkl} &= (1/6)[2(\chi^{ijkl} + \chi^{klij}) - (\chi^{iklj} + \chi^{ljik}) - (\chi^{iljk} + \chi^{jkil})], \tag{84a} \\
{}^{(Ax)}\chi^{ijkl} &= \chi^{[ijkl]} = \varphi\, e^{ijkl}, \tag{84b} \\
{}^{(Sk)}\chi^{ijkl} &= (1/2)\,(\chi^{ijkl} - \chi^{klij}). \tag{84c}
\end{align}
$$

Decomposition (84) is unique. The systematic study of skewonful cases started in 2002 (See, e.g., Hehl and Obukhov [7]). The Hehl-Obukhov-Rubilar skewon field (84c) can be represented as

$$ {}^{(Sk)}\chi^{ijkl} = e^{ijmk} S_m{}^l - e^{ijml} S_m{}^k, \qquad (85)$$

where $S_m{}^n$ is a traceless tensor with $S_m{}^m = 0$ [7]. Using a fiducial metric [e.g., the Lorentz metric (*h*-metric in the locally inertia frame), the *h*-metric or the symmetric part of an asymmetric metric] to raise/lower the indices, we can classify and resolve the skewon $S^{mn}$ into symmetric part (Type I; 9 degrees of freedom) and antisymmetric part (Type II; 6 degrees of freedom).

Media with constitutive tensor including skewon part has variety of wave-propagation structures and dispersion relations. We have studied systems with both core metric part (15) and skewon part in weak field/dilute medium situation. The various results are listed in Table 2. See [22, 33] for details.

Table 2. Various 1st-order and 2nd-order effects in wave propagation on media with the core-metric based constitutive tensors. $^{(P)}\chi^{(c)}$ is the extra contribution due to antisymmetric part of asymmetric metric to the core-metric principal part for canceling the skewon contribution to birefringence/amplification-dissipation [22,33].

| Constitutive tensor | Birefringence (in the geometric optics approximation) | Dissipation/ amplification | Spectroscopic distortion | Cosmic polarization rotation |
|---|---|---|---|---|
| Metric: ½ $(-h)^{1/2}[h^{ik} h^{jl} - h^{il} h^{kj}]$ | No | No | No | No |
| Metric + dilaton: ½ $(-h)^{1/2}[h^{ik} h^{jl} - h^{il} h^{kj}]\psi$ | No (to all orders in the field) | Yes (due to dilaton gradient) | No | No |
| Metric + axion: ½ $(-h)^{1/2}[h^{ik} h^{jl} - h^{il} h^{kj}] + \varphi e^{ijkl}$ | No (to all orders in the field) | No | No | Yes (due to axion gradient) |
| Metric + dilaton + axion: ½ $(-h)^{1/2}[h^{ik} h^{jl} - h^{il} h^{kj}]\psi + \varphi e^{ijkl}$ | No (to all orders in the field) | Yes (due to dilaton gradient) | No | Yes (due to axion gradient) |
| Metric + type I skewon | No to first order | Yes | Yes | No |
| Metric + type II skewon | No to first order; yes to 2nd order | No to first order and to 2nd order | No | No |
| Metric + $^{(P)}\chi^{(c)}$ + type II skewon | No to first order; no to 2nd order | No to first order and to 2nd order | No | No |
| Asymmetric metric induced: ½ $(-q)^{1/2}(q^{ik} q^{jl} - q^{il} q^{jk})$ | No (to all orders in the field) | No | No | Yes (due to axion gradient) |

## 7.1. Constitutive tensor from asymmetric metric

Eddington [80], Einstein and Straus [81], and Schrödinger [82, 83] considered asymmetric metric in their exploration of gravity theories. Just like we can build spacetime constitutive tensor from the (symmetric) metric as in metric theories of gravity, we can also build it from the asymmetric metric. Let $q^{ij}$ be the asymmetric metric as follows:

$$\chi^{ijkl} = \tfrac{1}{2} (-q)^{1/2}(q^{ik}q^{jl} - q^{il}q^{jk}), \tag{86}$$

with $q = \det^{-1}(^{(S)}q^{ij})$. When $q^{ij}$ is symmetric, this definition reduces to that of the metric theories of gravity. The constitutive law (86) was also put forward by Lindell and Wallen [84] as Q-medium. Resolving the asymmetric metric into symmetric part $^{(S)}q^{ij}$ and antisymmetric part $^{(A)}q^{ij}$:

$$q^{ij} = {}^{(S)}q^{ij} + {}^{(A)}q^{ij}, \text{ with } {}^{(S)}q^{ij} \equiv \tfrac{1}{2}(q^{ij} + q^{ji}) \text{ and } {}^{(A)}q^{ij} \equiv \tfrac{1}{2}(q^{ij} - q^{ji}), \tag{87}$$

we can decompose the constitutive tensor into the principal part $^{(P)}\chi^{ijkl}$, the axion part $^{(Ax)}\chi^{ijkl}$ and skewon part $^{(Sk)}\chi^{ijkl}$ as follows [33,85]:

$$\chi^{ijkl} = \tfrac{1}{2} (-q)^{1/2}(q^{ik}q^{jl} - q^{il}q^{jk}) = {}^{(P)}\chi^{ijkl} + {}^{(Ax)}\chi^{ijkl} + {}^{(Sk)}\chi^{ijkl}, \tag{88a}$$

with

$$^{(P)}\chi^{ijkl} \equiv \tfrac{1}{2}(-q)^{1/2}(^{(S)}q^{ik\,(S)}q^{jl} - {^{(S)}q^{il\,(S)}q^{jk}} + {^{(A)}q^{ik\,(A)}q^{jl}} - {^{(A)}q^{il\,(A)}q^{jk}} - 2^{(A)}q^{[ik\,(A)}q^{jl]}), \quad (88b)$$

$$^{(Ax)}\chi^{ijkl} \equiv (-q)^{1/2\,(A)}q^{[ik\,(A)}q^{jl]}, \quad (88c)$$

$$^{(Sk)}\chi^{ijkl} \equiv \tfrac{1}{2}(-q)^{1/2}(^{(A)}q^{ik\,(S)}q^{jl} - {^{(A)}q^{il\,(S)}q^{jk}} + {^{(S)}q^{ik\,(A)}q^{jl}} - {^{(S)}q^{il\,(A)}q^{jk}}). \quad (88d)$$

The axion part $^{(Ax)}\chi^{ijkl}$ only comes from the second order terms of $^{(A)}q^{il}$.

Using $^{(S)}q^{ij}$ to raise and its inverse to lower the indices, we have as equation (16) in [33]

$$S_{ij} = \tfrac{1}{4}\,\varepsilon_{ijmk}\,^{(A)}q^{mk};\ ^{(A)}q^{mk} = -\varepsilon^{mkij} S_{ij}, \quad (89)$$

where $\varepsilon_{ijmk}$ and $\varepsilon^{mkij}$ are respectively the completely antisymmetric covariant and contravariant tensors with $\varepsilon^{0123} = 1$ and $\varepsilon_{0123} = -1$ in local inertial frame. Thus the skewon field $S_{ij}$ from asymmetric metric $q^{ik}$ is antisymmetric and is of Type II with respect to $^{(S)}q^{ij}$.

*Dispersion relation in the geometrical optics limit.*

There are two ways to obtain the Tamm-Rubilar tensor density (51) for the dispersion relation (50). One way is by straightforward calculation; the other is by covariant method [85]. In the Appendix of the arXiv version of Ref. [22], we outline the straightforward calculation to obtain the Tamm-Rubilar tensor density $G^{ijkl}(\chi)$ for the asymmetric metric induced constitutive tensor:

$$G^{ijkl}(\chi) = (1/8)(-q)^{3/2}\det(q^{ij})\,q^{(ij}q^{kl)} = (1/8)(-q)^{3/2}\det(q^{ij})\,^{(S)}q^{(ij\,(S)}q^{kl)}. \quad (90)$$

Except for a scalar factor, (90) is the same as for metric-induced constitutive tensor with $^{(S)}q_{ij}$ replacing the metric $g_{ij}$ or $h_{ij}$. Therefore in the geometric optical approximation, there is no birefringence and the unique light cone is given by the metric $^{(S)}q_{ij}$.

*Constraints on asymmetric-metric induced constitutive tensor* [22]. Although the asymmetric-metric induced constitutive tensor leads to a Fresnel equation which is nonbirefringent, it contains an axionic part:

$$^{(Ax)}\chi^{ijkl} \equiv (-q)^{1/2\,(A)}q^{[ik\,(A)}q^{jl]} = \varphi\,e^{ijkl};\ \varphi \equiv (1/4!)\,e_{ijkl}(-q)^{1/2\,(A)}q^{[ik\,(A)}q^{jl]}, \quad (91)$$

which induces polarization rotation in wave propagation. Constraints on CPR and its fluctuation limit the axionic part and therefore also constrain the asymmetric metric. From Table 1, the mean of $\varphi$ ($\equiv (1/4!)\,e_{ijkl}(-q)^{1/2\,(A)}q^{[ik\,(A)}q^{jl]}$) is limited by observations on the cosmic polarization rotation to $< 0.02$ and its fluctuation to $< 0.02$ since the last scattering surface, and in turn constrains the antisymmetric metric of the spacetime for this degree of freedom. The antisymmetric metric has 6 degrees of freedom. Further study of the remaining 5 degrees of freedom experimentally to find either evidence or more constraints would be desired.

Theoretically, there are two issues: one is whether the asymmetric-metric induced constitutive tensors with additional axion piece are the most general nonbirefringent media in the lowest geometric optics limit; the other is what they play in the spacetime structure and in the cosmos.

## 8. How is the Lorentz-Minkowski Metric Coming about, Prospects and Discussions

Modern physics is strongly built on Lorentz-Minkowski metric. How can we "derive" this metric? We have the following prospects:

After the cosmological electoweak (vacuum) phase transition around 100 ps (particle energy scale about 100 GeV) from the Big Bang, high energy photons came out from their predecessors. At this time it was difficult to do measurement, although things might still evolve according to precise physical law – notably quantum electrodynamics and classical electrodynamics. When our Universe cooled down, precision metrology became possible. Metrological standards could be defined and implemented according to the fundamental physical laws.[23] The cosmic propagation according to WEP I (Weak Equivalence Principle I) for photons (nonbirefringence) in the framework of premetric classical electrodynamics of continuous media dictates that the spacetime constitutive tensor must be of core metric form with an axion (pseudoscalar) degree of freedom and a dilaton (scalar) degree of freedom. Propagation of pulsar pulses, radio galaxy signals and cosmological gamma ray bursts has verified this conclusion empirically down to $10^{-38}$, i.e. to $10^{-4} \times O([M_{Higgs}/M_{Planck}]^2)$. This is also the order that the generalized closure relations of electrodynamics are verified empirically. The axion and dilaton degrees of freedom are further constrained empirically in the present phase of the cosmos (Table 1). However, we should give a different thought to the axion and dilaton degrees of freedom in exploring spacetime and gravitation in the very early universe within 100 ps from the 'Big Bang'; we may need to look for imprints of new physics and new principles. Are there imprints from axions, dilatons or antisymmetric metric (skewons)? These would be clues to physical laws in very early universe. Testing WEP II will be a good way to decipher this. There is still a conformal degree of freedom in the the core metric. This conformal degree of freedom may be broken by the Higgs mass-induction. With this, the radiation-matter interaction fixes a unique metric. To test it, experiments with spin are important.

Related to these prospects, Hammond[86] searches for a relation betwwen stringy Maxwell charge and the magnetic dipole moment. Denisov, Ilyina and Sokolov[87] search for the nonlinear vacuum electrodynamical influence on the spacetime structure. Kruglov[88] calculates the universe acceleration due to nonlinear electromagnetic field. Stoica[89] expound expounds Kaluza theory with zero-length extra dimension. All these efforts may facilitate ways to explore the origins of gravity. Since axionic medium does exist, efforts to find corresponding media with dilatons and antisymmetric skewons will be waranteed. Since Ohm's law is not manifestly relativistic covariance, it would be nice to see a manifestly covariance form. Starke and Schober[90] review the explicit covariance of Ohm's law.

**Acknowledgement**

I would like to thank Science and Technology Commission of Shanghai Municipality (STCSM-14140502500) and Ministry of Science and Technology of China (MOST-2013YQ150829, MOST-2016YFF0101900) for supporting this work in part.